\documentclass[twocolumn, aps, prb]{revtex4}
\usepackage{graphicx}
\usepackage{amssymb}
\usepackage{amsmath}

\def\lsim{\lower -0.3ex \hbox{$<$} \kern -0.75em \lower 0.7ex \hbox{$\sim$}}
\def\gsim{\lower -0.3ex \hbox{$>$} \kern -0.75em \lower 0.7ex \hbox{$\sim$}}

\newcommand{\GVec}[1]{\mbox{\boldmath$#1$}}

\def\GVec#1{\mbox{\boldmath $#1$}}

\def\vare{\varepsilon}

\begin{document}

\title{Interlayer screening effect in
graphene multilayers with ABA and ABC stacking}
\author{Mikito Koshino}
\affiliation{
Department of Physics, Tokyo Institute of Technology\\
2-12-1 Ookayama, Meguro-ku, Tokyo 152-8551, Japan
}
\date{\today}

\begin{abstract}
We study the effect of perpendicular electric fields
on the band structures of ABA and ABC graphene multilayers,
and find that the electronic screening effect is significantly different
between them.
In ABA multilayers, the field produces a band overlap and gives
a linear screening, while
in ABC multilayers, in contrast, it opens an energy
gap in the surface-state band at low energy,
leading to a strong screening effect
essentially non-linear to the field amplitude.
The energy gap of a large ABC stack
sharply rises when the external field exceeds a certain critical value.
\end{abstract}

\maketitle

\section{Introduction}

Recent experimental realizations of atomically-thin graphene
systems \cite{novo04,novo05,zhang05} open up possibilities
of exploring their exotic electronic properties.
In multilayer films composed of more than two graphene layers,
the interlayer coupling strongly 
modifies the linear dispersion of monolayer graphene,
resulting in various electronic structures
depending on the number of layers, $N$.
\cite{novo06, Ohta06, Ohta07, McCa06, Koshino_and_Ando_2006a, 
Nils06, Guin06, Lati06, Lu06, Part06, Koshino_and_Ando_2007b, manes07,
Koshino_and_Ando_2008a, kosh09_ssc, kosh09_aba, avet09, guett08, crac09,avet09b}
The band structure can also be changed by applying a gate electric
field perpendicular to the layer,
through generating  an interlayer potential asymmetry.
In bilayer graphene, for example, 
an energy gap opens between the conduction and valence bands
in presence of gate electric field
\cite{McCa06,Lu06,Guin06,mcc06b,mcc07,min07,aoki07,gava09,avet09b}
and it was actually observed 
in transport \cite{castro07,oost08} and spectroscopic
measurements \cite{Ohta06,Ohta07,li09,zhang08,kuzmenko,zhang09,mak09}.

In nature, there are two known forms of bulk graphite called
ABA (AB, hexagonal, or Bernal) and ABC (rhombohedral) with different
stacking manners as shown in Fig.\ \ref{fig_atom}.
The ABA phase is thermodynamically stable and common,
while it is known that some portion of natural graphite
takes the ABC form. \cite{lipson42}
For ABA graphite, 
the effective mass model  describing the electronic property
was developed for the bulk system
\cite{Wall47,Slon58,McCl56,McCl57,McCl60,Dres65,dressel02},
and also for few-layer systems.
\cite{McCa06,Koshino_and_Ando_2006a,Nils06,Guin06,Lu06,Part06,
Koshino_and_Ando_2008a, kosh09_ssc,Koshino_and_Ando_2007b,manes07,kosh09_aba}.
The energy dispersion of the multilayer graphenes
includes the subbands analog to monolayer or the bilayer
graphene, \cite{Guin06,Part06} and the Hamiltonian is actually 
decomposed into independent subsystems 
effectively identical to monolayer or bilayer.
\cite{Koshino_and_Ando_2007b,Koshino_and_Ando_2008a}
The ABC graphite has a quite different electronic structure from ABA's 
\cite{haering58,mcclure69,
Guin06,lu06abc,Lati06,aoki07,lu07abc,min08,arovas08,kosh09_abc}. 
In particular, the low-energy band of a finite ABC multilayer
are given by the surface states localized 
at outer-most layers, \cite{Guin06,manes07}
and the interlayer potential asymmetry opens an energy gap in those
bands. \cite{aoki07,lu07abc,kosh09_abc}
This is in sharp contrast with ABA multilayers
where potential asymmetry causes a band overlapping. \cite{aoki07,kosh09_aba}

In considering the interlayer potential asymmetry 
induced by an external electric field, 
it is essential to take into account 
screening effect, as done in bilayer graphene,\cite{mcc06b,mcc07,min07} 
and ABA multilayers \cite{guin_scr,kosh09_aba,avet09b}.
Experimentally, the interlayer screening effect 
in the gate electric field 
was probed in thin graphite films. \cite{Ohta07,Miya08,Lee09,Sui09}
Here we calculate the self-consistent band structure 
of ABA and ABC multilayers with various $N$'s 
in the presence of perpendicular electric field.
For ABA multilayers, we show that
the electric field generally produces band overlapping,
and the screening is shown to be linear to the field amplitude.
In ABC multilayers, on the other hand,
the low-energy surface band causes a strong non-linear screening effect 
through opening an energy gap.
The paper is organized as follows: 
we present the effective mass models for ABA and ABC multilayers
in Sec.\ \ref{sec_band}, 
and compute the band structure 
including the self-consistent screening effect in Sec.\ \ref{sec_scr}.
The conclusion is given in Sec. \ \ref{sec_concl}

\section{Effective Hamiltonian and band structure}
\label{sec_band}

\subsection{ABA multilayers}

We first consider a multilayer graphene with ABA stacking,
composed of $N$ layers of a graphene layers.
We label $A$ and $B$ on $i$-th layer as $A_i$ and $B_i$.
In ABA stacking, the sites
$B_1, A_2, B_3, A_4 \cdots$ are arranged along vertical columns
normal to the layer plane, while the rest sites
$A_1, B_2, A_3, B_4 \cdots$ are above or below 
the center of hexagons in the neighboring layers,
as shown in Fig.\ \ref{fig_atom} (a).
The system is described by a {\bf k}$\cdot${\bf p} Hamiltonian
based on three-dimensional (3D) graphite model.
\cite{Wall47,Slon58,McCl56,McCl57,McCl60,Dres65,dressel02}
As the simplest approximation, 
we include parameter $\gamma_0$
describing the nearest neighbor coupling within each layer,
and $\gamma_1$ for the coupling of the interlayer vertical bonds.
The band parameters were experimentally estimated
in the bulk ABA graphite, for example \cite{dressel02} 
as  $\gamma_0=3.16$ eV and $\gamma_1 = 0.39$ eV,
which we will use in the following calculations.
The lattice constant of honeycomb lattice
(distance between nearest $A$ atoms) is given by $a = 0.246$ nm,
and the inter-layer spacing $d = 0.334$ nm. 

The low-energy spectrum is given by the states 
in the vicinity of $K$ and $K'$ points in the Brillouin zone.
Let $|A_j\rangle$ and $|B_j\rangle$ be the Bloch functions at the $K$
point, corresponding to the $A$ and $B$ sublattices, respectively, of
layer $j$. If the basis is taken as $|A_1\rangle,|B_1\rangle$;
$|A_2\rangle,|B_2\rangle$; $\cdots$; $|A_N\rangle,|B_N\rangle$, the
Hamiltonian around the $K$ point 
is given by \cite{Guin06,Part06,Lu06,Koshino_and_Ando_2007b}
\begin{eqnarray}
 {\cal H}_{\rm ABA} =
\begin{pmatrix}
 H_1 & V  \\
 V^{\dagger} & H_2 & V^{\dagger}  \\
 & V & H_3 & V \\
 &   & \ddots & \ddots & \ddots & 
\end{pmatrix},
\label{eq_H_AB} 
\end{eqnarray}
and
\begin{eqnarray}
&& H_j = 
\begin{pmatrix}
 U_j & v p_- \\ v p_+ & U_j
\end{pmatrix}, 
\quad
V = 
\begin{pmatrix}
0 & 0 \\ \gamma_1 & 0
\end{pmatrix},
\label{eq_H_suppl} 
\end{eqnarray}
where  $U_j$ is the electrostatic potential at $j$th layer,
and we defined $p_\pm = p_x \pm i p_y$ with
$\GVec{p} = -i\hbar \nabla$.
$v$ is the band velocity of monolayer graphene
given by $v = \sqrt{3} a \gamma_0/2\hbar$.
The effective Hamiltonian for another valley, $K'$,
is obtained by interchanging $p_+$ and $p_-$. \cite{McCa06}

When $U_j=0$, Hamiltonian (\ref{eq_H_AB})
can be decomposed into subsystems identical to
bilayer or monolayer graphenes
with a basis appropriately chosen. \cite{Koshino_and_Ando_2007b}
The subsystems are labeled by an index $m$ which ranges as
\begin{eqnarray}
 m = 
\left\{
\begin{array}{l}
1,3,5, \cdots , N-1, \quad N = {\rm even} \\
0,2,4, \cdots , N-1, \quad N = {\rm odd}
\end{array}
\right.
\label{eq_m}
\end{eqnarray}
The eigenenergies at $U_j=0$ are given \cite{Guin06,Koshino_and_Ando_2007b}
for $m=0$ as $\vare_{m=0,s}^{\rm ABA}(p) = s v p$,
and for $m \neq 0$ as
\begin{eqnarray}
 \vare_{m,\mu,s}^{\rm ABA}(p) = 
s\left[
\mu \gamma_1 \cos \kappa_m
+
\sqrt{
(\gamma_1 \cos \kappa_m)^2
+ (v p)^2}
\right],
\label{eq_ene_ABA}
\end{eqnarray}
where $p=\sqrt{p_x^2+p_y^2}$, $\mu = \pm$, $s =\pm$,
and
\begin{eqnarray}
\kappa_m = \frac{\pi}{2}-\frac{m\pi}{2(N+1)}.
\end{eqnarray}

$m=0$ only exists in odd-layer graphene and gives
an energy band identical to monolayer graphene.
Other $m$'s are bilayer-type band structures
where $\mu=-$ gives a pair of electron ($s=+$) and hole bands ($s=-$) touching
at zero energy, and $\mu=+$ another pair repelled away
by $\pm 2 \gamma_1 \cos \kappa_m$.
The dispersion around $k = 0$
is approximately quadratic with the effective mass \cite{McCa06}
\begin{equation}
m^* =
\frac{\gamma_1}{v^2} \cos \kappa_m
\label{eq_mstar},
\end{equation}
giving the density of states at zero energy,
 $\rho_m = g_v g_s m^*/(2\pi\hbar)$,
with $g_v=2$ and $g_s=2$ are valley ($K,K'$) and spin degeneracies,
respectively.

The quantity $\kappa_m$ corresponds to the wave number $k_z$
in the layer stacking direction ($z$-direction) via $\kappa_m = k_z d$.
\cite{Guin06,Koshino_and_Ando_2007b}
The wave function of subband $m$
is indeed a standing wave in $z$-direction with wave number $\kappa_m$. 
The total density of states per layer, 
$\bar{\rho} = (1/N) \sum_m \rho_m$, 
approximates in large $N$ limit,
\begin{equation}
\bar{\rho} =
g_v g_s \frac{\gamma_1}{2\pi^2\hbar^2 v^2},
\label{eq_rho_tot}
\end{equation}
where $\sum_m$ is replaced with integration in $\kappa$.


\subsection{ABC multilayers}

\begin{figure}
\begin{center}
 \leavevmode\includegraphics[width=70mm]{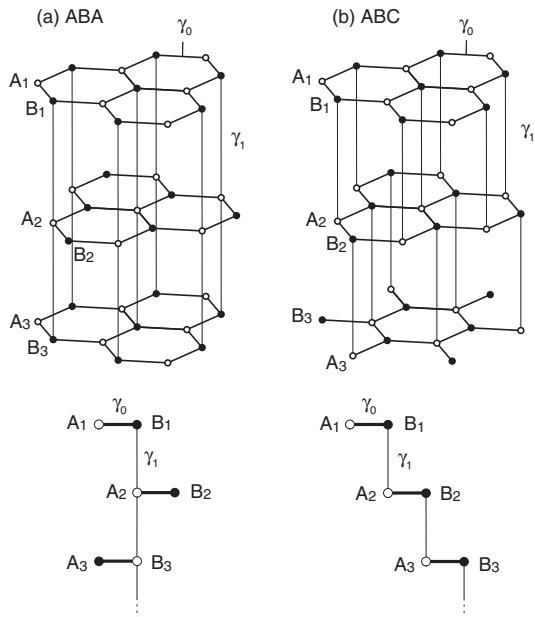}
\end{center}
\caption{
Atomic structures of multilayer graphenes with
(a) ABA (Bernal) stacking and
(b) ABC (rhombohedral) stacking
} 
\label{fig_atom}
\end{figure}

The ABC multilayer have a different arrangement
shown in Fig.\ \ref{fig_atom} (b), where
vertical bonds couple the pairs $(B_j, A_{j+1})$
for $j=1,2,\cdots,N-1$.
We use the same notation $\gamma_0$ and 
$\gamma_1$ as in ABA graphite, 
for the nearest intralayer and interlayer coupling, respectively.
Although the band parameters 
are not identical between ABA and ABC graphites, 
we refer to the values of ABA
in the following numerical calculations, assuming that the
corresponding coupling parameters have similar values. \cite{mcclure69}
Hamiltonian around the $K$ point 
can be written as \cite{mcclure69,Guin06,lu06abc,arovas08}
\begin{eqnarray}
 {\cal H}_{\rm ABC} =
\begin{pmatrix}
 H_1 & V  \\
 V^{\dagger} & H_2 & V  \\
 & V^{\dagger} & H_3 & V \\
  &   & \ddots & \ddots & \ddots & 
\end{pmatrix},
\label{eq_H_ABC}
\end{eqnarray}
with the same matrices defined in Eq.\ (\ref{eq_H_suppl}).
When $U_j$=0, the eigenenergies are given by
\begin{eqnarray}
 \vare_{n,s}^{\rm ABC}(p) = 
s \sqrt{(v p)^2 + \gamma_1^2
+ 2\gamma_1 vp\cos{\varphi_n}},
\label{eq_ene_ABC}
\end{eqnarray}
with $s=\pm$ and $\varphi_n$ $(n=1,2,\cdots,N)$ being solutions of
\begin{equation}
 vp \sin (N+1)\varphi + \gamma_1 \sin N\varphi = 0.
\label{eq_phi}
\end{equation}
The corresponding wavefunction is
$|\psi\rangle = \psi(A_1)|A_1\rangle+\psi(B_1)|B_1\rangle+\cdots$
with
\begin{eqnarray}
\begin{pmatrix}
\psi(A_j) \\
\psi(B_j)
\end{pmatrix}
=C
\begin{pmatrix}
e^{i\theta(j-1)}\, \sin(N+1-j)\varphi_n \\
s e^{i\theta j}\, \sin j\varphi_n
\end{pmatrix},
\label{eq_wave}
\end{eqnarray}
where $\theta =\arctan p_y/p_x$
and $C$ is a normalization factor.
In the bulk limit, $\varphi_n$ corresponds to the wavenumber
along the layer stacking ($z$) direction.
Actually, Eq.\ (\ref{eq_phi}) is obtained by 
imposing a condition that
a standing wave in $z$-direction,
composed by bulk wavefunctions,
becomes zero at fictitious sites $B_0$ and $A_{N+1}$
out of the system.

\begin{figure*}
\begin{center}
 \leavevmode\includegraphics[width=150mm]{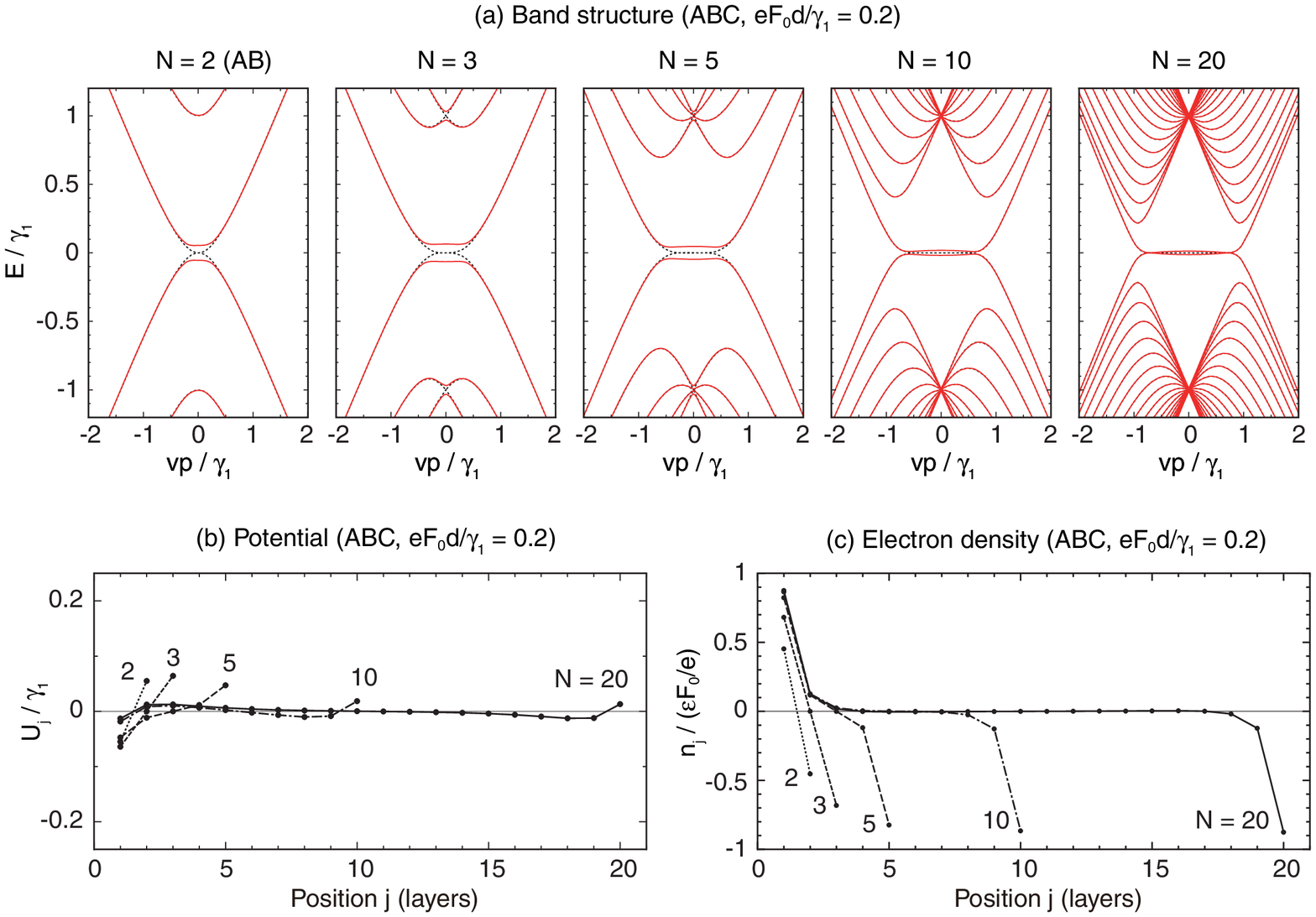}
\end{center}
\caption{
(a) Self-consistent band structures of 
ABC (rhombohedral) multilayer graphenes with several layer number $N$'s,
at external field $eF_0 d /\gamma_1 = 0.2$ (red, solid) 
and 0 (black, dotted).
(b) Potential distribution and (c) electron density
of ABC multilayers with several $N$'s at $eF_0 d /\gamma_1 = 0.2$.
}
\label{fig_band_abc}
\end{figure*}

Equation (\ref{eq_phi}) has $N$ solutions of $\varphi$
giving independent eigenstates.
All of $\varphi_n$ are real when $vp>\gamma_1N/(N+1)$, 
while only one becomes complex when $vp < \gamma_1 N/(N+1)$,
which corresponds to the evanescent mode in the bulk.
In $vp \ll \gamma_1$, the complex branch approximates
$e^{i\varphi} \approx - vp /\gamma_1$,
giving the dispersion 
\begin{equation}
\vare \approx s \gamma_1 \left(vp /\gamma_1\right)^N,
\label{eq_en_edge}
\end{equation}
with $s=\pm$.
These are the only bands
which appear at $\vare=0$ and dominate the low-energy physics.
The corresponding wavefunction is
\begin{eqnarray}
\begin{pmatrix}
\psi(A_j) \\
\psi(B_j)
\end{pmatrix}
\approx C
\begin{pmatrix}
e^{i\theta(j-1)}\, (-vp /\gamma_1)^{N+1-j} \\
s e^{i\theta j}\, (-vp /\gamma_1)^j
\label{eq_wav_edge}
\end{pmatrix}.
\end{eqnarray}
The wave amplitude becomes largest on the top and bottom layers
and decays exponentially inside, and thus is
regarded as a surface state \cite{Guin06}.
The wave function is exactly localized at the sites
$A_1$ and $B_N$ at $p=0$, and as $p$ increases, the decay length 
increases as $-1/\log (vp/\gamma_1)$ in units of interlayer spacing $d$.
In Fig.\ \ref{fig_band_abc}, we plot 
the band structures of ABC graphenes with $N=2,3,5,10$ and 20,
where the results of $U_j=0$ are indicated as black dotted curves. 
The surface states of Eq.\ (\ref{eq_en_edge}) are shown as
a pair of electron and hole bands touching at $\vare=0$, 
which become flatter as $N$ increases.
The bilayer graphene (AB) can be regarded
as $N=2$ of ABA family and also that of ABC family,
and indeed, equally described either of Eqs. (\ref{eq_ene_ABA})
or (\ref{eq_ene_ABC}).

When we consider the low-energy physics around
zero energy, it is convenient to use the effective Hamiltonian
reduced to the basis  $|A_1\rangle,|B_N\rangle$. \cite{McCa06,manes07,min08}
In presence of $U_j$, it reads
\begin{eqnarray}
\mathcal{H}_{\rm ABC}^{\rm (eff)} =
\begin{pmatrix}
U_1 & \gamma_1 (v p_-/\gamma_1)^N \\
\gamma_1 (v p_+/\gamma_1)^N & U_N
\end{pmatrix}.
\label{eq_H_ABC_eff}
\end{eqnarray}
This approximation is valid when $vp/\gamma_1 \ll 1$,
i.e., the actual wave function, Eq.\ (\ref{eq_wav_edge}),
is well localized to $A_1$ or $B_N$.
When we set the origin of potential as $U_1+U_N=0$,
the eigenenergy 
is given by
\begin{equation}
 \vare_{s,p} = s \sqrt{\gamma_1^2 (v p/\gamma_1)^{2N} +(\Delta U/2)^2}, 
\label{eq_ene_ABC_eff}
\end{equation}
where $\Delta U = U_1 - U_N$ and $s=\pm$. 
The potential difference $\Delta U$
opens an energy gap between the valence and conduction bands.



\begin{figure*}
\begin{center}
 \leavevmode\includegraphics[width=150mm]{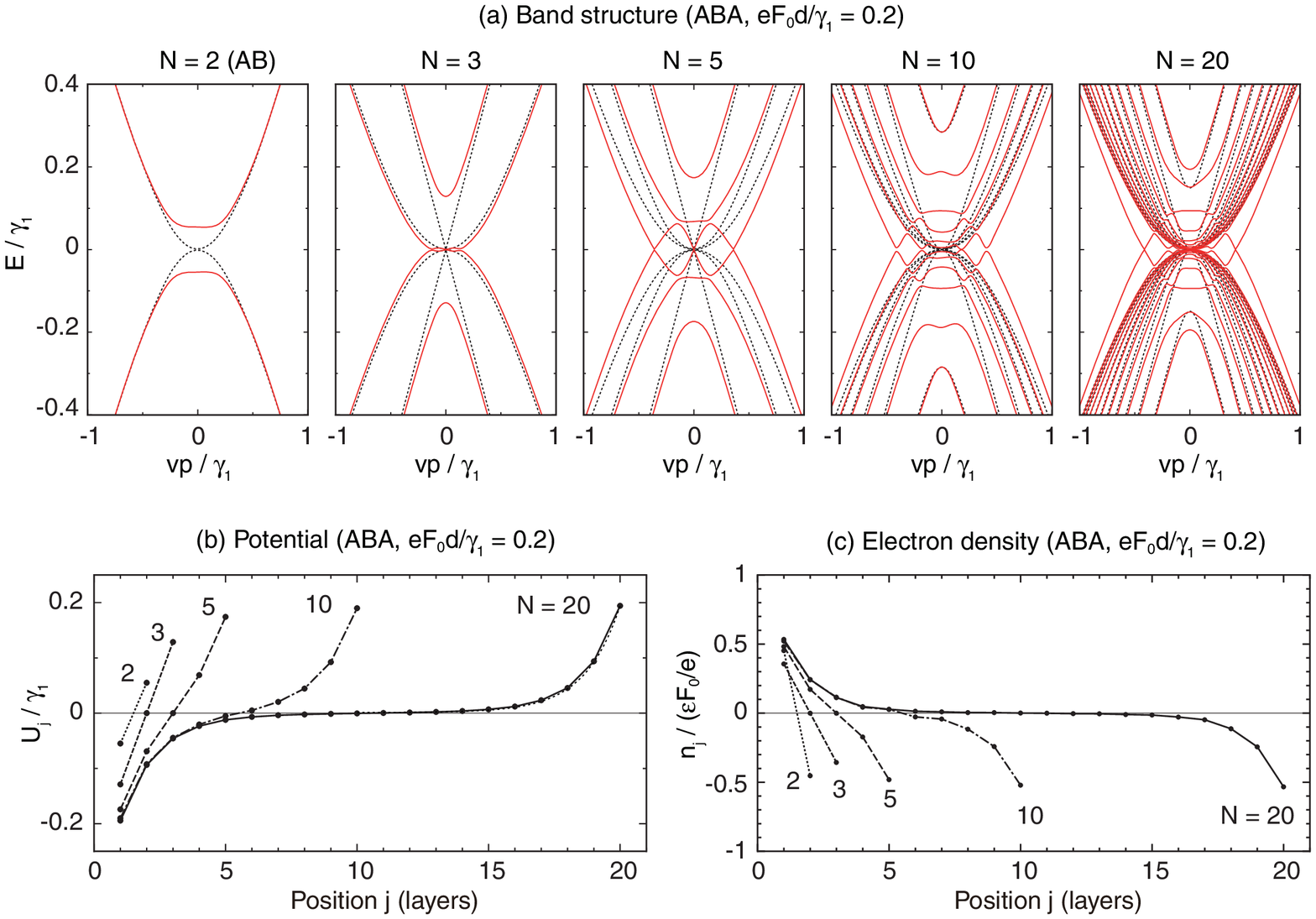}
\end{center}
\caption{
(a) Self-consistent band structures of 
ABA (Bernal) multilayer graphenes with several layer number $N$'s,
at external field $eF_0 d /\gamma_1 = 0.2$ (red, solid) 
and 0 (black, dotted).
(b) Potential distribution and (c) electron density
of ABA multilayers with several $N$'s at $eF_0 d /\gamma_1 = 0.2$.}
\label{fig_band_ab}
\end{figure*}

\begin{figure}
\begin{center}
 \leavevmode\includegraphics[width=60mm]{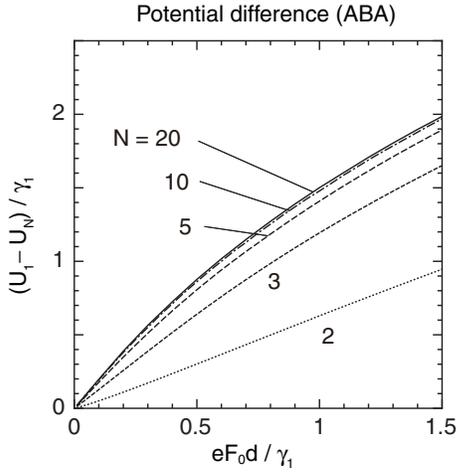}
\end{center}
\caption{
Potential difference between outermost layers
as a function of external field,
in ABA multilayer graphenes with several $N$'s. 
} 
\label{fig_potdiff_ab}
\end{figure}


\begin{figure}
\begin{center}
 \leavevmode\includegraphics[width=60mm]{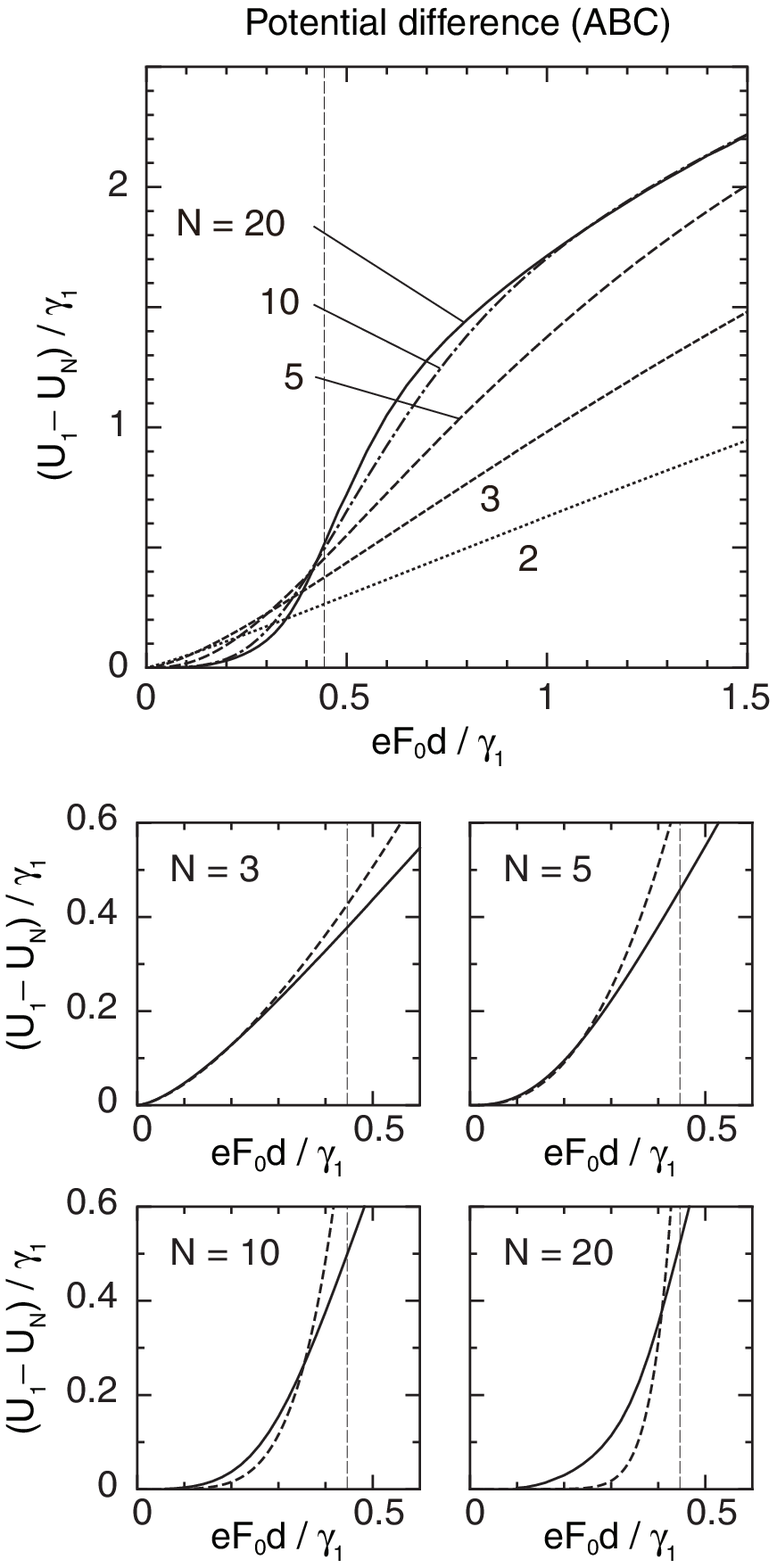}
\end{center}
\caption{
Potential difference between outermost layers
of ABC multilayers as a function of external field.
Vertical broken line indicates the critical field $F_c$.
Lower four panels compare the same results (solid curves) to
the approximate expressions
of Eq.\ (\ref{eq_delta_u}) (dashed).
} 
\label{fig_potdiff_abc}
\end{figure}

\section{Screening effects}
\label{sec_scr}

\subsection{Self-consistent treatment of screening effect}

We compute the band structure 
of ABA or ABC multilayer graphenes
in presence of gate electric field taking account of the 
screening effect.
We consider undoped free-standing multilayer graphenes
with an external electric field $F_0$ applied to the perpendicular
direction.
This situation can be realized in an experimental set up
with an external top and bottom gates electrodes 
which are held at the opposite gate voltages
with respect to the graphene. \cite{kosh09_aba}


The potential at each layer, $U_j (j=1,2,\cdots,N)$
should be determined self-consistently.
If a set of $U_j$ is given, we can compute the band structure
using the Hamiltonian Eq.~(\ref{eq_H_AB}) for ABA
or Eq.~(\ref{eq_H_ABC}) for ABC multilayers.
Then we determine the Fermi energy so that the total density is
equal to $n_{\rm tot}$ ($=0$ in the present case),
and calculate the electron density at each layer, 
$n_j (j=1,2,\cdots,N)$, from the occupied eigenstates.
For screening effect, we consider the multilayer as parallel plates 
with zero thickness and respective electron densities $n_j$.
The electric field between $j$th and $(j+1)$th layers
is then given by
\begin{equation}
 F_{(j,j+1)} = F_0 + 
 \frac{e}{2\vare} 
\left[
\sum_{j'=1}^j  n_{j'} 
-
\sum_{j'=j+1}^N n_{j'}
\right].
\label{eq_Fj}
\end{equation}
Here $\varepsilon$ is the permittivity of the 
interlayer spaces without the screening effect of $\pi$-band electrons,
and we set $\vare = 2$ in the following calculations.
Eq.\ (\ref{eq_Fj}) immediately gives 
a new set of the electrostatic potential $U_j$,
which should be identical to the initial $U_j$.
To find the self-consistent solution,
we employ an iterative numerical approach,
where we start with $U_j = eF_0 [j-(N+1)/2]$ as initial values
and iterate the process until $U_j$'s converge.

\subsection{ABA multilayers}

In Fig.\ \ref{fig_band_ab} (a), 
solid red curves show the self-consistent
band structures of ABA multilayers with several $N$'s,
in presence of the external field $e F_0 d = 0.2\gamma_1$.
The original band structures at $F_0 = 0$ are also shown as
dotted black curves.
In $N \geq 3$, we see that the lowest electron band is pulled down
and the highest hole band is lifted up,
making a band overlap around zero energy,
as was previously 
recognized in the case of $N=3$ and 4. \cite{aoki07,kosh09_aba} 
The energy width of overlap becomes almost constant in $N \gsim 10$.

Figures \ref{fig_band_ab}(b) and 
\ref{fig_band_ab}(c) show the corresponding potential
distribution $U_j$ and electron density $n_j$, respectively,
at the same external field $e F_0 d = 0.2\gamma_1$.
In $N \geq 10$, 
we observe that the electric field (i.e., gradient in $U_j$) 
is screened within a few layers from the surface, 
leaving a triangular potential pocket at each end.
The potential decay near the edge
is almost identical between $N=10$ and 20.
The overlapping bands observed in Fig.\ \ref{fig_band_ab} (a)
are actually the bound states trapped at either of pockets;
the states of the lowest electron and the highest hole bands
are indeed localized at the potential minimum (left end)
and maximum (right), respectively.
Since $E_F$ is zero,
those bands are populated by 
electrons or holes, 
contributing to the most part of the screening field.
A smooth decay observed in the electron density
appears different from Ref. \cite{guin_scr}, 
which finds a charge oscillation with every second layer.
We presume that this is due to the contribution from
the intraband excitation, which was dropped in numerical calculations 
for neutral systems in Ref. \cite{guin_scr}.

The typical screening length $\lambda_s$ 
(penetrating depth of electric field) 
can be roughly estimated by Thomas-Fermi approximation. \cite{guin_scr}
In this treatment, the potential decay on the surface is expressed as
$U(z) \propto e^{-z/\lambda_s}$ with
$\lambda_s = (e^2 \rho_{\rm 3D}/\vare)^{-1/2}$,
where $\rho_{\rm 3D}$ is the 
three dimensional density of states at the Fermi energy.
For graphene, if we substitute $\rho_{\rm 3D} = \bar\rho /d$
with $\bar\rho$ of Eq.\ (\ref{eq_rho_tot}),
we obtain \cite{guin_scr}
\begin{equation}
 \lambda_s =  \left(g_v g_s
\frac{\gamma_1}{2\pi^2\hbar^2 v^2}
\frac{e^2}{\vare d}
\right)^{-1/2}.
\label{eq_lambda_s}
\end{equation}
Using the parameters above, we get 
$\lambda_s \sim 1.3 d \approx 0.43$nm.
In Fig.\ \ref{fig_band_ab} (b), 
we plot an exponential curve with decay length $\lambda_s$
in Eq.\ (\ref{eq_lambda_s}) 
as a dotted curve to fit with the right half of the curve of $N=20$,
which shows a fairly nice agreement.
The depth of potential depth, or $|U(z=0)|$,
is roughly estimated as $e F_0 \lambda_s$,
which determines the order of the energy width
in band overlapping.

Figure \ref{fig_potdiff_ab} displays
the potential difference $\Delta U = U_1-U_N$ as a function of
the external field $F_0$.
$\Delta U$ rises almost linearly in increasing $F_0$,
except for a slight sub-linear components in large $F_0$.
This is consistent with Thomas-Fermi approximation,
since it gives linear screening in a weak external field.

\subsection{ABC multilayers}

The screening property of ABC multilayers is quite different from
that of ABA, as the density of states diverges 
at $\vare=0$ due to the flat band of the surface states.
Before numerical calculations with full band model,
we present an analytical approach using the effective $2\times 2$
Hamiltonian of Eq.\ (\ref{eq_H_ABC_eff}) valid in low energies.
The potential difference $\Delta U$ between the top and bottom layers
opens an energy gap between the valence and conduction bands,
and thus only the lower band $(s=-)$ is occupied
when $n_{\rm tot} =0$.
The density difference between the top and bottom layers,
$\delta n = n(A_1) - n(B_N)$, is calculated as
\begin{eqnarray}
 \delta n &=&  \frac{g_v g_s}{L^2}\sum_{p} 
|\psi_{-,p}(A_1)|^2 - |\psi_{-,p}(B_N)|^2 \nonumber\\
&=&
\frac{g_vg_s}{2\pi} 
\left(\frac{\gamma_1}{\hbar v}\right)^2
\left(\frac{\Delta U}{2\gamma_1}\right)^{2/N}
f_N,
\label{eq_delta_n}
\end{eqnarray}
where $(\psi_{-,p}(A_1),\psi_{-,p}(B_N))$ is the eigenvector
of Eq.\ (\ref{eq_H_ABC_eff}) for $s=-$ band,
and
\begin{equation}
 f_N = \int_0^\infty \frac{t\,dt}{\sqrt{t^{2N}+1}}
= \frac{
\Gamma\left(\frac{1}{2}-\frac{1}{N}\right)
\Gamma\left(1+\frac{1}{N}\right)
}{2\sqrt{\pi}},
\label{eq_fn}
\end{equation}
with $\Gamma(x)$ is the gamma function.

The density imbalance $\delta n$ causes the screening field
$F_{\rm ind} = - e \delta n/(2\vare)$ opposed to
the external field $F_0$, resulting in the total potential difference
$\Delta U = e (F_0 + F_{\rm ind}) (N-1) d $.
Together with Eq.\ (\ref{eq_delta_n}), we obtain
the self-consistent equation for $\Delta U$,
\begin{equation} 
\Delta U = 
e (N-1) d 
\left[
F_0  - \frac{e}{2\vare}
\frac{g_vg_s}{2\pi} 
\left(\frac{\gamma_1}{\hbar v}\right)^2
\left(\frac{\Delta U}{2\gamma_1}\right)^{2/N}
f_N
\right].
\label{eq_sc}
\end{equation}
In $N \geq 3$,
$\Delta U$ is negligible compared to $\Delta U^{2/N}$
when $\Delta U$ is small enough.
Then the equation is solved approximately as
\begin{equation} 
\Delta U \approx
2\gamma_1 F_0^{N/2} 
\left[
 \frac{e}{2\vare}
\frac{g_vg_s}{2\pi} 
\left(\frac{\gamma_1}{\hbar v}\right)^2
f_N
\right]^{-N/2},
\label{eq_delta_u}
\end{equation}
which is essentially non-linear in $F_0$.
In large-$N$ limit, we have $f_N \approx 1/2$ and thus 
$\Delta U \approx 2\gamma_1 (F_0/F_c)^{N/2}$,
where $F_c = e n_c/(2\vare)$ is a characteristic field
with an associated electron density
\begin{equation}
n_c =
\frac{g_vg_s}{4\pi} 
\left(\frac{\gamma_1}{\hbar v}\right)^2 
\approx 1.2 \times 10^{13} {\rm cm}^{-2}.
\end{equation}
In increasing $F_0$, 
$\Delta U$ rapidly increases when the external field exceeds $F_c$.

$n_c$ is the electron density 
accommodated in the flat-band region in large $N$ limit ($vp/\gamma_1 < 1$),
i.e., the number of surface states.
The field is completely screened in $F_0<F_c$,
because the surface states are able to supply
positive and negative charge 
to opposite surfaces to cancel the external field.
The screening collapses at $F_c$,
when the density required for canceling
exceeds the surface states population $n_c$.

$N=2$ (AB) is an exceptional in that 
the integration in Eq.\ (\ref{eq_fn}) diverges logarithmically,
giving infinite $\delta n$.
Actually this is an artifact of the reduced $2\times 2$ model,
due to the incorrect
contributions from large $p$ where the reduced Hamiltonian 
is not accurate. We can remove this
by introducing a momentum cut-off $p_c \sim \gamma_1/v$,
and get $f_{N=2} \sim (1/2)\log({\gamma_1/\Delta U})$.
When we neglect the logarithmic dependence of $f_N$,
$\Delta U$ becomes linear in $F_0$
in accordance with Eq.\ (\ref{eq_sc}).
The logarithmic factor gives a weak singularity
at $\Delta U = 0$.

Now we numerically calculate 
the self-consistent band structure of ABC multilayers
using the full Hamiltonian Eq.\ (\ref{eq_H_ABC}).
Figure \ref{fig_band_abc} (a) shows
the results at $e F_0 d = 0.2\gamma_1$ (red, solid)
and 0 (black, dotted).
In presence of the external field, an energy gap opens at low-energy 
as expected. The gap width becomes smaller in $N>5$
in increasing $N$, suggesting a strong screening effect
in large stacks.
Figures \ref{fig_band_abc}(b) and \ref{fig_band_abc}(c) show
the corresponding potential distribution $U_j$
and the electron density $n_j$, respectively, 
at the same field $e F_0 d = 0.2\gamma_1$.
At $N=20$, the potential is almost flat inside,
as the external field is mostly screened 
by the electric charge on surface states 
localized at the outermost layers.
This is in contrast with ABA multilayers, 
where an external field always penetrates inside with a few-layer thickness.

Figure \ref{fig_potdiff_abc} shows the
plots of the potential difference $\Delta U = U_1-U_N$ as a function of
the external field $F_0$.
We actually observe non-linear behavior
expected in the analytical argument,
where $\Delta U$ rapidly increases at $F_0 \sim F_c$
(shown as a dashed vertical line).
Lower panels in Fig.\ \ref{fig_potdiff_abc} compare the numerical results 
(solid) to the analytical expression Eq.\ (\ref{eq_delta_u}) (dashed).
We have nice agreements for $N \leq 5$ in small $F_0$,
while the approximation becomes worse in large stack of $N\geq 10$.
In large $N$'s, the low-energy band almost reaches $vp/\gamma_1 \sim 1$,
where the wave function deeply penetrates into the bulk
in accordance with Eq.\ (\ref{eq_wav_edge}).
The finite penetration length makes the screening less effective,
compared to the previous $2\times 2$ model 
assuming the wave functions perfectly localized on the surface layers.
As a result, numerical curves  at large $N$'s
rise less sharply than the analytical ones as observed
in Fig.\ \ref{fig_potdiff_abc}.
The wave penetration to the bulk is also responsible for
Mexican hat structure, \cite{Guin06}
or narrowing of the gap around $vp/\gamma_1 \sim 1$
observed in Fig.\ \ref{fig_band_abc} (a).
There the actual energy splitting becomes smaller than $\Delta U$
because the wave function is not perfectly localized 
at surface layers.

\begin{figure}
\begin{center}
 \leavevmode\includegraphics[width=60mm]{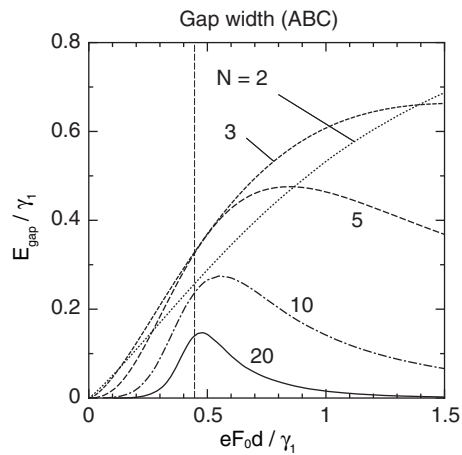}
\end{center}
\caption{
Energy gap width in the self-consistent band structures 
as function of external field,
for ABC multilayers with several $N$'s.
Vertical broken line indicates the critical field $F_c$.
}
\label{fig_gap}
\end{figure}

The width of the energy gap is an important quantity
which can be detected experimentally.
Figure \ref{fig_gap} shows
the gap width against the external field
in the self-consistent band structures of ABC multilayers.
In $N \leq 5$, the band bottom is approximately flat and
the gap width therefore approximates $\Delta U$ (the splitting
at $p=0$), and actually rises in proportional to $F_0^{N/2}$.
In large stacks of $N \geq 10$, 
the energy gap becomes maximum around $F_0 \sim F_c$,
and is suppressed in greater $F_0$'s, 
due to the gap narrowing at $vp/\gamma_1 \sim 1$.

\section{Conclusion}
\label{sec_concl}

We studied electronic band structures
of ABA and ABC graphene multilayers in the presence of 
an perpendicular electric field, including the screening effect.
In ABA multilayers, 
the electric field produces band overlapping
accompanying a linear screening 
well described by the Thomas-Fermi approximation.
In ABC multilayers, in contrast, 
the surface state bands dominating low energies
cause a strong non-linear screening effect
through opening an energy gap.


While in the present model we only include the primary 
parameters $\gamma_0$ and $\gamma_1$ in our model,
the extra band parameters corresponding to the
further hopping generally affect the band structure
of multilayer graphenes. 
\cite{Wall47,Slon58,McCl57,McCl60,Dres65,dressel02}
In ABC graphenes, it was shown that the extra parameters 
gives a fine structure to the surface band,
of which energy scale is expected to be
of the order of 10 meV. \cite{kosh09_abc}
We expect that the screening property would be
influenced by those effects when the external 
potential is as small as those energy scales.
As another remark, the electron-electron interaction 
other than the screening effect
may create non-trivial ground states in a flat band such as in 
ABC multilayers, while we leave those problems for future works.

\section*{Acknowledgments}

The author thanks E. McCann and T. Ando for helpful discussions.
This work was supported in part by Grant-in-Aid for Scientific Research
on Priority Area ``Carbon Nanotube Nanoelectronics'' and by Grant-in-Aid
for Scientific Research from Ministry of Education, Culture, Sports,
Science and Technology Japan.


\end{document}